\documentclass[11pt,a4paper]{article}
\usepackage[utf8]{inputenc}
\usepackage[english]{babel}
\usepackage{amsmath}
\usepackage{amsfonts}
\usepackage{amssymb}
\usepackage{lmodern}
\usepackage[left=2cm,right=2cm,top=2cm,bottom=2cm]{geometry}
\usepackage{tensor}
\usepackage{braket}

\allowdisplaybreaks  

\usepackage{tikz}
\usetikzlibrary{positioning,arrows}
\usetikzlibrary{decorations.pathmorphing}
\usetikzlibrary{decorations.markings}

\newcommand{\dd}[1]{d#1\ }

\newcommand{\ie}{i.e.\
}
\newcommand{\eg}{e.g.\
}

\newcommand{\st}{stress-energy tensor}

\newcommand{\cO}{{\cal{O}}}
\newcommand{\cA}{{\cal{A}}}

\title{Conformal Anomaly for Non-Conformal Scalar Fields}
\author{Lorenzo Casarin\footnote{lorenzo.casarin@aei.mpg.de}, \;  Hadi Godazgar\footnote{hadi.godazgar@aei.mpg.de} \; and Hermann Nicolai\footnote{hermann.nicolai@aei.mpg.de} \\Max-Planck-Institut f\"ur Gravitationsphysik (Albert-Einstein-Institut), \\
Am M\"uhlenberg 1, D-14476 Potsdam, Germany. }
\date{ September 18, 2018 }

\begin{document}

\tikzset{
particle/.style={thick,draw=black, postaction={decorate},
    decoration={markings,mark=at position .60 with {\arrow[black]{triangle 45}}}},
aparticle/.style={thick,draw=black, postaction={decorate},
    decoration={markings,mark=at position .5 with {\arrow[black]{triangle 45,rotated}}}},
gluon/.style={decorate, draw=black,
    decoration={coil,aspect=0}}
 }

\maketitle
\abstract

\noindent
We give a general definition of the conformal anomaly for theories that are not classically Weyl 
invariant and show that this definition yields a quantity that is both finite and local. As an
example we study the conformal anomaly for a non-minimally coupled massless scalar 
and show that our definition coincides with results obtained using the heat kernel method. 

\section{Introduction}

Conformal anomalies have been the focus of much study since the 1970s  \cite{CapDuff,Duff1,DDI,Dowker,Brown,Tsao,BD,FT,FT1,FT2,DS,Duff2,BGVZ,BvanN,ST,Tseytlin}.
One aspect, however, that has not received so much attention concerns 
the significance of conformal anomalies in theories that are {\em not} classically
conformal (or Weyl) invariant. One main reason for the interest in  this question 
is the fact that the most interesting cancellations of conformal anomalies occur in 
{\em non}-conformal supergravities for $N\geq 5$ \cite{MN1} (whereas conformal
supergravities stop at $N=4$, and cancellations require very special matter couplings \cite{FT1}).
The full significance of these cancellations has not been fully understood so far;
moreover there is a pending issue concerning the dependence of the $a$ 
and $c$ coefficients on the gauge choice for the external gravitational field for
fields of spin $\geq \frac32$ that remains to be resolved. In this paper we focus 
on the investigation of the conformal anomaly and its significance for the 
specific example of the non-conformal scalar field. For this we follow
a recent recalculation of the anomaly for spin-$\frac12$ fields  \cite{GN}, which
is based on an evaluation of Feynman diagrams largely analogous to (though
much more involved than) the textbook derivation of the axial anomaly in gauge theories.
The spin-0 field coupled to a background gravitational field represents the simplest
example in which to consider non-conformal deformations, and has the added advantage 
that there is a tunable parameter $\xi$, such that for one special value of $\xi$ the 
theory becomes conformal. 

For any theory, whether conformal or not, we adopt the following {\em general definition}
of the conformal anomaly
\begin{equation}\label{A}
\mathcal{A} (x) : = \lim_{\varepsilon\rightarrow 0} 
\Big[g^{(4)\mu\nu} \big\langle T_{\mu\nu}(x) \big\rangle -
	 \big\langle    g^{\mu\nu} T_{\mu\nu}(x) \big\rangle \Big]\,.
\end{equation}
where $\varepsilon \equiv \frac12(d-4)$ is the regularization parameter in dimensional
regularization, and the superscript on $g^{\mu\nu}$  indicates the dimension in which the 
trace is to be performed (we do not yet specify this dimension in the second term, because
there is a choice which does not affect the physically significant part of the anomaly,
as we will explain below). Here the second term removes the classical violation of conformal 
invariance, reflected in a non-vanishing trace of the classical {\st}    -- the difference 
between the quantum trace and the expectation value of the classical trace is what 
produces the conformal anomaly.  We show in section \ref{sec:anomaly} that this 
definition yields a result that is always finite and local for any theory, including theories 
that are not classically Weyl invariant, and hence to which one would not normally
associate a Weyl anomaly. Furthermore we will see that in the non-conformal case
the two terms on the r.h.s. of (\ref{A}) by themselves do not produce a meaningful 
answer because they separately exhibit divergences and non-local terms which 
only cancel in the difference: this is the reason the difference must be taken {\em before}
removing the regulator. In section~2 below we present a general argument why 
this is always true. Furthermore, in the case of a non-minimally coupled scalar 
we show that the coefficients of ${\bf E}_4$ (Euler number) and $C^2$ in the anomaly do 
not depend on the value of $\xi$, whereas the coefficient of $\Box R$ does 
depend on $\xi$ (but the $a$ and $c$ coefficients may start to depend on various
couplings in higher loop orders in the presence of interactions, as there appears to be 
no analog of the Adler-Bardeen theorem for the conformal anomaly).

For the non-minimally coupled scalar, our conformal anomaly agrees with the heat 
kernel coefficient $a_2$ up to the coefficient of the scheme-dependent 
contribution $\Box R$ \cite{BD}, giving an interpretation for the heat kernel 
coefficient in the non-conformal case.
Although our calculation therefore mostly recovers known results, our  derivation
differs from previous 
ones and exhibits several new features. One of these is that, for non-conformal theories, 
the anomaly as defined in (\ref{A}) need {\em not} satisfy the Wess-Zumino (WZ) 
consistency condition
\begin{equation}
\frac{\delta\big(\sqrt{-g}\cA(x)\big)}{\delta\sigma(y)} \,=\,
\frac{\delta\big(\sqrt{-g}\cA(y)\big)}{\delta\sigma(x)}
\end{equation}
where $\sigma \equiv \log(-g)$ is the conformal factor. 
As a result on the r.h.s. of (\ref{A}) for non-conformal theories
there will appear in addition to the usual ${\bf E}_4\,$, $C^2$  and $\Box R$ anomalies 
(which do satisfy the WZ condition) extra terms proportional to $R^2$ (which does not 
satisfy the WZ condition). A further new result of this paper is the explicit structure of 
the pole terms which has not been given in the literature to the best of our knowledge, 
but which can alternatively be derived from heat kernel methods as we will show.

While the properties of a quantum field theory at criticality are well-understood because 
of Weyl symmetry, or conformal symmetry in the flat space limit, it is less understood 
how these properties are modified away from criticality. For example, it is known that there is a monotonically decreasing function that interpolates between a UV fixed point and an 
IR fixed point \cite{Komargodski:2011vj}. By studying the trace of the stress tensor away from criticality we hope to identify properties of the trace that may be preserved even away from 
the point, albeit, in this case, with a deformation that requires a non-trivial metric background.

\section{Conformal anomaly for non-Weyl invariant theories} \label{sec:anomaly}

Before entering into the details of the spin zero case, we would like to present a general
argument why (\ref{A}) always produces a finite and local result, provided all divergences are local. As we said above, we wish to define the analogue of the conformal anomaly for theories that are not necessarily classically Weyl invariant. In the general case, the trace of the expectation value of the stress 
tensor (first term on the r.h.s. of (\ref{A})) will be both divergent and non-local. 
Even if we renormalise the theory in order to remove the divergence we will only be guaranteed a local expression when the theory is classically Weyl invariant, the expression thus being the anomaly. If we however take definition \eqref{A}, which reduces to the standard anomaly for theories that are Weyl invariant, we can show that the expression will be both finite and local. 

The expectation value of the operator $T_{\mu\nu}(x)$, in a theory regularised with dimensional regularisation, reads
\begin{equation}\label{aaa}
\braket{T_{\mu\nu}(x)} = \frac{P_{\mu\nu}(x)}{\varepsilon}  + F_{\mu\nu}(x)
\end{equation}
where $P_{\mu\nu}$ and $F_{\mu\nu}$ are the pole and the finite part of the expansion of the expectation value. In general such expectation value is divergent (and requires renormalisation to yield a finite result). Here we show that, provided the pole $P_{\mu\nu}$ is local, the quantity (\ref{A})  
is finite and local. We stress that this is the case in any local quantum field theory where the 
calculation is performed to $n$-loops provided all the divergences from $(n-1)$ loops 
have been subtracted; in particular this is true for $n=1$, that is the case of interest here. 
For the conformal anomaly, as defined in equation \eqref{A}, to be local at higher loop order 
we must therefore use the action renormalised up to the previous loop order. 
The first term on the r.h.s. of (\ref{A}) is to be computed considering the four-dimensional 
trace (\ie $g^{\mu\nu}g_{\mu\nu} = 4$) of \eqref{aaa} {\em after} computing the regularised
expectation value of $T_{\mu\nu}$. For the second we take the classical trace
{\em before} computing the expectation value. Here we have the choice of taking
the trace in $d$ or in four dimensions. For the $d$-dimensional trace we note the identity
\begin{equation}\label{dvs4}
g^{(d)\mu\nu} \big\langle T_{\mu\nu} \big\rangle \,=\,
\big\langle g^{(d)\mu\nu}  T_{\mu\nu} \big\rangle
\end{equation}
whence the two operations of taking the trace and taking the expectation value commute
with this choice. This is no longer the case if the trace is taken in four dimensions. If the 
trace inside the bracket is taken in $d$ dimensions we thus arrive at the alternative
formula
\begin{equation}\label{AA}
\mathcal{A} (x) : = \lim_{\varepsilon\rightarrow 0} 
\Big[ \big( g^{(4)\mu\nu} - g^{(d)\mu\nu} \big) \big\langle T_{\mu\nu}(x) \big\rangle \Big].
\end{equation}
For the special example of the scalar field studied in this paper we will find that the difference 
$\big\langle  ( g^{(d)\mu\nu} - g^{(4)\mu\nu}) T_{\mu\nu}(x) \big\rangle$ is
proportional to $\Box R$, and can thus be absorbed into a local counterterm.
Hence the choice of dimension in the second term on the r.h.s. of (\ref{A})
has no intrinsic physical significance, at least for the case at hand.
The above formula also shows why the WZ consistency condition fails if the 
second term on the r.h.s. of (\ref{AA}) does not vanish: there is then no functional
differential operator $\delta/\delta\tilde\sigma(x)$ to reproduce the r.h.s.
by acting on some regularized effective functional.

From these observations it follows that
\begin{align}
\label{aac}
g^{(4)\, \mu\nu}\braket{T_{\mu\nu}(x)} 
& =  \frac{g^{(4)\, \mu\nu} P_{\mu\nu}(x)}{\varepsilon}  + g^{(4)\, \mu\nu} F_{\mu\nu}(x) 
=  \frac{ P( x,4)}{\varepsilon} + F(x, 4)
\\[2mm]
\label{aad}
\braket{ g^{(4-2\varepsilon)\, \mu\nu} \,  T_{\mu\nu}(x)}  
& =  \frac{g^{(4-2\varepsilon)\, \mu\nu} P_{\mu\nu}(x)}{\varepsilon}  + g^{(4-2\varepsilon)\, \mu\nu} F_{\mu\nu}(x) 
=  \frac{ P(x, 4-2\varepsilon)}{\varepsilon}  + F(x,4-2\varepsilon)
\end{align}
making use of (\ref{dvs4}) and defining 
$P(x, d):= g^{(d)\, \mu\nu} P_{\mu\nu}(x) $, namely the second argument of $P$ is the trace of $g$ (similarly for $F$) or the dimension of spacetime. Expanding \eqref{aad} in powers of $\varepsilon$  yields for $\mathcal{A} (x) $ the expression
\begin{equation}\label{aae}
\mathcal{A} (x) = 2 P'(x,4) + O(\varepsilon)
\end{equation}
where the $'$ indicate derivative with respect to second argument. This discussion also shows that the terms contributing to \eqref{aae} are only those with an explicit factor of $g_{\mu\nu}(x)$, and not other tensors for which the difference between a contraction in different dimensions vanishes (\eg $g^{\mu\nu}R_{\mu\nu} = R$ both for $D=4$ and for $D= 4 - 2\varepsilon$). We will verify this explicitly in the case of the non-minimally coupled scalar.

\section{Scalar field}

We start with the action for  a real scalar in $d$ dimensions\footnote{Throughout this
paper we use the mostly plus signature \(\eta_{\mu\nu} = \mathrm{diag}({}-{}+{}+{}+{})\).}
\begin{equation}
	S
= 
	- \frac{1}{2} \int \dd{^d x} \sqrt{-g} \
		\phi
		\left(
			-\square  + \xi R
		\right)
		\phi
		\label{action}
\end{equation}
with the associated  stress-energy tensor 
\begin{align}\label{Tmunu-definition}
	T_{\mu\nu}
& =
	\frac{2}{\sqrt{-g}} g_{\mu\alpha} g_{\nu\beta} \frac{\delta S}{\delta g_{\alpha\beta}}
\\
& =
	\partial_\mu \phi \partial_\nu \phi
	- \frac{1}{2} g_{\mu\nu} \partial_\alpha \phi  \partial^\alpha \phi
	+ \xi \phi^2 \left( R_{\mu\nu } - \frac{1}{2} g_{\mu\nu} R \right)
	- \xi \left( \nabla_\mu \partial_\nu \phi^2 - g_{\mu\nu} \nabla^\alpha \partial_\alpha \phi^2 \right).
\end{align}
By virtue of the equation of motion ($\Box \phi = \xi R \phi$) this trace is covariantly 
conserved, $\nabla^\mu T_{\mu\nu} = 0$, for any $\xi$. In $d$ dimensions,
this action is furthermore conformally invariant if and only if  $\xi= \xi_d$ with
\begin{equation}\label{xid}
\xi_d \equiv \frac{d-2}{4(d-1)}.
\end{equation}
Accordingly, the trace of the stress-energy tensor vanishes on-shell for $\xi = \xi_d$, since 
\begin{equation}\label{Tmunu-trace}
	g^{\mu\nu} T_{\mu\nu} = 2 (d-1) (\xi - \xi_d)
	\left( \partial_\alpha \phi \partial^\alpha \phi   + \xi R\phi^2 \right) 
	= (d-1) (\xi - \xi_d) \square \big( \phi^2 \big).
\end{equation}

For the perturbative determination of the anomaly (\ref{A}) we follow the same procedure 
as in ~\cite{GN} and expand the metric in the usual way
\begin{equation}
	g_{\mu\nu} = \eta_{\mu\nu} + h_{\mu\nu}.
\end{equation}
with ensuing expansions of the action and the {\st} in powers of $h_{\mu\nu}$,
\begin{eqnarray}
S &=&  S^{(0)} + S^{(1)} + S^{(2)} + \ldots \\[1mm]
T_{\mu\nu} &=& T_{\mu\nu}^{(0)} + T_{\mu\nu}^{(1)} 
            + T_{\mu\nu}^{(2)} + \ldots
\end{eqnarray}
where the superscript $(n)$ corresponds to the collection of all term of $\cO(h^n)$.
The quantity to be computed is then
\begin{equation}
\big\langle T_{\mu\nu} (x) \big\rangle =
\left\langle \big( T_{\mu\nu}^{(0)} (x)   + T_{\mu\nu}^{(1)} (x) + \cdots \big)
    e^{i (S^{(1)} + S^{(2)} + \cdots)} \right\rangle_0
\end{equation}
where $\langle\cdots\rangle_0$ refers to the free scalar expectation value
(below we will often drop the subscript $0$ when it is obvious what is meant).
The evaluation of the Feynman diagrams resulting from this expansion is completely
analogous to the calculation performed in \cite{GN} to which we refer for further 
technical details.

\section{Computations at $\cO(h)$}

The computations at $\cO(h)$ are straightforward, and are only included here for
completeness. At first order in the metric perturbation the expectation value of the 
stress tensor is
\begin{equation}
	\left\langle T_{\mu\nu} (x) \right\rangle
=
	 i \left\langle T^{(0)}_{\mu\nu} (x)  S^{(1)}  \right\rangle_0  \; + \; \cO(h^2).
\end{equation}
We write
\begin{align}
	i \left\langle T^{(0)}_{\mu\nu} (x)  S^{(1)}  \right\rangle_0
\label{tmunu-twopint}
& =
	- i \int \dd{^dy} \int \frac{d^d p}{(2\pi)^d} e^{-ip(x-y)} T_{\mu\nu\alpha\beta}(p) h^{\alpha\beta}(y),
\end{align}
where $ T_{\mu\nu\alpha\beta}(p)$ is the two-point function of stress tensor,
\begin{equation}\label{2-pt}
\begin{tikzpicture}[baseline={([yshift=-.5ex]current bounding box.center)},thick]
\draw (1,0) node[right]{$\mu \nu$};
        \draw (2,0) circle [radius=0.17];
        \draw (2-0.12, - 0.12) -- (2+0.12, 0.12);
        \draw (2+0.12,  -0.12) -- (2-0.12,  0.12);
\draw [particle] (3.2,0) arc [radius=0.6, start angle=0, end angle= 180];
\draw [particle] (2,0) arc [radius=0.6, start angle=180, end angle= 360];
\draw (4.2,0.7) node[right]{$ $};
\draw (4.2,0) node[right]{$\alpha \beta$};
\draw (2.5, 0.7) node[above]{$k -p$};
\draw (2.4, - 0.65) node[below]{$k$};
\draw (3.8, 0.25) node[above]{$p$};
\draw [<-] (3.6,0.2) -- (4,0.2);
\draw[gluon](4.2,0)--(3.2,0); 
\end{tikzpicture}
\hspace{1em}
= 
	T_{\mu\nu\alpha\beta} ( p )
=
	\int \frac{d{^d k}}{(2\pi)^{d}} \frac{1}{k^2 (k-p)^2} 
	V_{ \mu \nu } ( k - p , - k ) \
	V_{ \alpha \beta  } ( k , p - k ) 
\end{equation}
with
\begin{equation}\label{Vmunu}
\begin{tikzpicture}[baseline={([yshift=-.5ex]current bounding box.center)}, thick]
\draw[gluon](-1.2,0)--(0,0);
\draw[particle](1,1,0)--(0,0); 
\draw[particle](1,-1,0)--(0,0);
\draw (-1.2,0) node[left]{$\tau\sigma$};
\draw (-0.80,0.25) node[above]{$-k-\ell$};
\draw [<-] (-0.45,0.25) -- (-0.85,0.25);
\draw (0.4,0.55) node[above]{$k$};
\draw (0.6,-0.5) node[above]{$\ell$};
\end{tikzpicture}
=
\begin{tikzpicture}[baseline={([yshift=-.5ex]current bounding box.center)}, thick]
        \draw (0,0) circle [radius=0.17];
        \draw (-0.12, - 0.12) -- (0.12, 0.12);
        \draw (0.12,  -0.12) -- (-0.12,  0.12);
\draw[particle](1,1,0)--(0,0); 
\draw[particle](1,-1,0)--(0,0);
\draw (-0.15,0) node[left]{$\tau\sigma$};
\draw (0.4,0.55) node[above]{$k$};
\draw (0.6,-0.5) node[above]{$\ell$};
\end{tikzpicture}
=
V_{\tau\sigma} (k,\ell) \,\equiv\, \frac12 (k \cdot \ell) \eta_{\tau\sigma} - k_{(\tau} \ell_{\sigma)}
               + \xi\Big( (k+\ell)_\tau (k+\ell)_\sigma - \eta_{\tau\sigma} (k+\ell)^2 \Big)
\end{equation}
Conservation of the stress tensor at $\cO(h)$ follows directly from 
\begin{equation}
p^{\mu} T_{\mu\nu\alpha\beta} ( p )=0,
\end{equation}
which itself is a consequence of the vanishing of tadpole integrals.

It is now straightforward to take the trace of the expectation value of the \st
\begin{equation}
	g^{\mu\nu} \big\langle T_{\mu\nu}(x) \big\rangle
=
	- i \eta^{\mu\nu} \int \dd{^4y} \int \frac{d^d p}{(2\pi)^d} e^{-ip(x-y)} T_{\mu\nu\alpha\beta}(p) h^{\alpha\beta}(y)
	+ \cO(h^2).
\end{equation}
We are interested in the result {\em for arbitrary} $\xi$. After some calculation, and taking 
the four-dimensional  trace of the regularised integral we obtain, up to higher 
powers of $\varepsilon$ 
\begin{align}\label{anomaly}
		\eta^{(4)\mu\nu} T_{\mu\nu\alpha\beta} (p)
& =
- \frac{i p^2 \left( p_\alpha p_\beta -\eta_{\alpha\beta} p^2  \right)}{(4\pi)^2} 
	\left[	
		\frac{\left( 6 \xi - 1 \right)^2}{12} \left(\frac{1}{ \varepsilon}+    2 -  \gamma_E- \log \frac{p^2}{4\pi \mu^2}  \right) 
		 - \frac{1}{15}  \left( \frac{ 11 }{ 12 } - 5\xi \right)  
	\right].
\end{align}
For non-conformal values of $\xi\neq \frac16$, this trace exhibits a pole as well as a 
non-local contribution $\propto \log (p^2/\mu^2)$  ($\mu^2$ is the usual regularisation 
scale). To remove these terms we next evaluate the expectation value of the regularised 
on-shell trace of the energy-momentum tensor  at order $\cO(h)$, \ie
\begin{equation}\label{adc}
	\big\langle g^{(d) \mu\nu} T_{\mu\nu} (x) \big\rangle\big|_{\cO(h)} \, =\,
	i  \Braket{ (d-1)(\xi-\xi_d) \Box( \phi^2) \, S^{(1)} }
=
	- i \int \dd{^dy} \int \frac{d^d p}{(2\pi)^d} e^{-ip(x-y)} \tau_{\alpha\beta}(p) h^{\alpha\beta}(y);
\end{equation}
where we kept the factor $(d-1)(\xi-\xi_d)$ inside the expectation value to indicate that it has to be expanded as $d=4-2\varepsilon$; $\tau_{\alpha \beta} (p) $ is given by the expression
\begin{equation}\label{tmumu}
	\tau_{\alpha\beta} (p)  \, \equiv \, 
	(d-1)(\xi-\xi_d) p^2 \int \frac{d^{d} k}{(2\pi)^{d}} 
		\frac1{k^2(k-p)^2}   V_{\alpha\beta} ( p - k ,  k ) .
\end{equation}
The evaluation of \eqref{tmumu} is straightforward, and leads to 
\begin{equation}
	\tau_{\alpha\beta} (p) 
\, = \,	- \frac{i p^2\left(
			p_\alpha p_\beta 
			- p^2 \eta_{\alpha\beta}
		\right) }{(4\pi)^2} 		
		\left[
			 \frac{\left( 6 \xi - 1 \right)^2}{12} \left(\frac{1}{\varepsilon} +   2 -  \gamma_E- \log \frac{p^2}{4\pi \mu^2}  \right) 
			- \frac{\left(
				6 \xi - 1
			\right) (3\xi-1)}{9} 
		\right].
\end{equation}
We see that both the pole and the non-local term match precisely 
with \eqref{anomaly} to produce a finite and local result. We also notice that the subtraction 
alters the coefficient of $\square R$, so in the limit $\varepsilon\rightarrow 0$ we end up with 
\begin{equation}\label{anxiD}
	\mathcal{A}_\xi
=
	\lim_{\varepsilon\rightarrow 0} \Big[ g^{\mu\nu} \big\langle T_{\mu\nu}(x) \big\rangle_\xi -
	 \langle    g^{(4-2\varepsilon) \mu\nu} T_{\mu\nu}(x) \rangle_\xi \Big]
=
	\frac{1}{ 180 ( 4 \pi )^2 } (1-10(1-6\xi)^2)  \square R.
\end{equation}

In removing the classical trace, we could also use dimensional regularisation by dimensional reduction whereby we treat the contractions over momenta (or derivatives in position space) as $d$-dimensional but traces as 4-dimensional.
The result in this case reads 
\begin{equation}\label{anxi4}
	\mathcal{A}^{(BD)}_\xi
\,=\,
	g^{\mu\nu} \big\langle T_{\mu\nu}(x) \big\rangle-
	 \langle g^{(4) \ \mu\nu} T_{\mu\nu}(x) \rangle
=
	\frac{1}{ 30 ( 4 \pi )^2 } (1-5\xi)  \square R.
\end{equation}
The new coefficient matches with that given by Birrell and Davies \cite[p.\ 179]{BD}.
We see that the different prescription affects the coefficient of the $\square R$ contribution, which is a scheme-dependent contribution and can in any case be tuned to any desired value by choice of a suitable
$R^2$ counterterm, whence this coefficient has no intrinsic significance. This is in marked
contrast to the coefficients of the ${\bf E}_4$ and $C^2$ anomalies at $\cO(h^2)$
which exhibit no such prescription dependence. This is the reason why
the nice trick that allows the $c$-coefficient to be determined from the $\cO(h)$ computation~\cite{Duff1} (see also~\cite{GN}) no longer works for non-conformal theories. Consequently for $\xi\neq \frac16$
the determination of the $a$ and $c$ coefficients requires a calculation at $\cO(h^2)$.

Let us also consider the tensor structure of the pole of $\braket{T_{\mu\nu}}$ at
first order in $h$ of the expectation value of the stress  energy tensor as computed through \eqref{tmunu-twopint}. Given that the expression must be local, generally covariant and must have dimension $+4$, this restricts it to the form
\begin{equation}\label{pole1}
\braket{T_{\mu\nu}} =  \frac1{(4\pi)^2\varepsilon}\Big[
a_1\ g_{\mu\nu} \square R
+ a_2\ \nabla_\mu \nabla_\nu R
+ a_3\ \square R_{\mu\nu}
\Big]
+
\cO(h^2).
\end{equation}
Using the first order expansions for the Ricci tensor and Ricci scalar, 
we can match the expansion term by term and we get 
\begin{align}
a_1  =  \frac{ -3 + 40 \xi - 120 \xi^2}{120},   \qquad
a_2  =  \frac{1 - 10 \xi + 30 \xi^2}{30}, \qquad
a_3  =  - \frac{1}{60}.
\end{align}
As a check, we can trace over $\mu\nu$ indices, and indeed we obtain, to first order in $h$
\begin{equation}
g^{\mu\nu}\braket{T_{\mu\nu}} = \frac1{(4\pi)^2}
\frac{4 a_1 + a_2 + a_3}{\varepsilon} \square R 
=  -\frac{ (1-6\xi)^2}{12 (4\pi)^2 \varepsilon}\square   R
\end{equation}
which matches with the pole of \eqref{anomaly}.

Furthermore, using the arguments of section \ref{sec:anomaly}, we can see that the anomaly, $\mathcal{A}_{\xi}$, should only depend on the coefficient $a_1$ in \eqref{pole1}, namely
\begin{equation}
 \mathcal{A}_{\xi} = \frac1{(4\pi)^2} \, 2 a_1,
\end{equation}
which indeed agrees with equation \eqref{anxiD}.

\section{Computations at  $\cO(h^2)$}
The computation is considerably more involved at second order in $h$,
but works along similar lines to those in ~\cite{GN}; for this reason we here display only the salient
results.~\footnote{Full details of the computation will be provided in the forthcoming
thesis by one of the authors (L. Casarin).}
At second order we have
\begin{equation}\label{gen}
	\left\langle T_{\mu\nu} (x) \right\rangle\Big|_{\cO(h^2)}
\,=\,
	 i \left\langle T^{(0)}_{\mu\nu} (x)  S^{(2)}  \right\rangle_0
	 -
	 \frac{1}{2} \left\langle T^{(0)}_{\mu\nu} (x)  S^{(1)}  S^{(1)}  \right\rangle_0	 
	 +
	 i \left\langle T^{(1)}_{\mu\nu} (x)  S^{(1)}  \right\rangle_0.
\end{equation}
We write
\begin{align} \label{diags}
	i \left\langle T^{(0)}_{\mu\nu} (x)  S^{(2)}  \right\rangle_0
& =
	- i \int \dd{^dy}\dd{^dz} \int \frac{d^d p}{(2\pi)^d}\frac{d^d q}{(2\pi)^d} e^{i p(x-y)}   e^{i q(z-y)}  
	h^{\alpha\beta}(y)  h^{\rho \sigma}(z)
	\ T^{[2]}_{\mu\nu\alpha\beta\rho\sigma}(p,q) ,
\\[2mm]
	- \frac{1}{2} \left\langle T^{(0)}_{\mu\nu} (x)  S^{(1)}  S^{(1)}  \right\rangle_0
& =
	- i \int \dd{^dy} \dd{^dz} \int \frac{d^d p}{(2\pi)^d} \frac{d^d q}{(2\pi)^d} e^{i p(x-y)}   e^{i q(z-y)}  
	h^{\alpha\beta}(y)  h^{\rho \sigma}(z)
	\ T^{[3]}_{\mu\nu\alpha\beta\rho\sigma}(p,q)  ,
\\[2mm]
	 i \left\langle T^{(1)}_{\mu\nu} (x)  S^{(1)}  \right\rangle_0
& =
	- i \int \dd{^dy}\dd{^dz} \int \frac{d^d p}{(2\pi)^d} \frac{d^d q}{(2\pi)^d} e^{i p(x-y)}   e^{i q(z-y)}
	h^{\alpha\beta  }(y)  h^{ \rho \sigma }(z)\
		T^{[4]}_{\mu\nu\alpha\beta\rho\sigma}(p+q,-q) .
\end{align}
In the last integral, we have rewritten $h(x)$ as the inverse Fourier transform of its Fourier transform and shifted the integration variables as \( (p,q) \rightarrow (p+q,-q)  \) to make the exponential factors uniform. 
The functions above read
\begin{align}
	T^{[2]}_{\mu\nu\alpha\beta\rho\sigma}(p,q) \nonumber
& = \notag 
\begin{tikzpicture}[baseline={([yshift=-.5ex]current bounding box.center)},thick]
\draw (1,0) node[right]{$ \mu \nu$};
        \draw (2,0) circle [radius=0.17];
        \draw (2-0.12, - 0.12) -- (2+0.12, 0.12);
        \draw (2+0.12,  -0.12) -- (2-0.12,  0.12);
\draw [particle] (2,0) arc [radius=0.6, start angle=180, delta angle= -180];
\draw [particle] (3.2,0) arc [radius=0.6, start angle=0, delta angle= -180];
\draw (4.2,0.7) node[right]{$\alpha \beta $};
\draw (4.2,-0.7) node[right]{$\rho \sigma$};
\draw (2.3, 0.7) node[above]{$ k - p $};
\draw (2.4, - 0.6) node[below]{$k$};
\draw[gluon](4.2,0.7,0)--(2+0.6+0.6,0); 
\draw[gluon](4.2,-0.7,0)--(2+0.6+0.6,0);
\draw (3.7, -0.7-0.05) node[below]{$ - q $};
\draw [->] (3.8, -0.7) -- (3.5,-0.5);
\draw (3.7, 0.6) node[above]{$ p + q$ };
\draw [->] (3.8, 0.7) -- (3.5,0.5);
\end{tikzpicture}
\\
& =
	\int \frac{d{^d k}}{(2\pi)^{d}} \frac{1}{k^2 (k-p)^2} 
	V_{ \mu \nu } ( k - p , - k ) \
	W_{ \alpha \beta  \rho\sigma } ( k-p,-k , p+q,-q ) ,
\\ \nonumber
\\ \notag
 	T^{[3]}_{\mu\nu\alpha\beta\rho\sigma}(p,q) 
& =       \nonumber
\begin{tikzpicture}[baseline={([yshift=-.5ex]current bounding box.center)},thick]
\draw (1,0) node[right]{$ \mu \nu$};
        \draw (2,0) circle [radius=0.17];
        \draw (2-0.12, - 0.12) -- (2+0.12, 0.12);
        \draw (2+0.12,  -0.12) -- (2-0.12,  0.12);
\draw [particle] (2,0) arc [radius=0.6, start angle=180, delta angle= -120];
\draw [particle] (2.9,0.52) arc [radius=0.6, start angle=60, delta angle= -120];
\draw [particle] (2.9,-0.52) arc [radius=0.6, start angle=300, delta angle= -120];
\draw (4.2,0.7) node[right]{$\alpha \beta $};
\draw (4.2,-0.7) node[right]{$\rho \sigma$};
\draw (2.2, 0.55) node[above]{$k-p $};
\draw (2.4, - 0.6) node[below]{$k$};
\draw (3.35, 0) node[right]{$k-q$};
\draw[gluon](4.2,0.7,0)--(2.9,0.52); 
\draw[gluon](4.2,-0.7,0)--(2.9,-0.52);
\draw (3.7, -0.7-0.05) node[below]{$-q$};
\draw [->] (3.85, -0.8-0.05) -- (3.4,-0.693-0.1);
\draw (3.7, 0.7+0.05) node[above]{$p+q$};
\draw [->] (3.85, 0.8+0.05) -- (3.4,0.693+0.1);
\end{tikzpicture}
\\
& =
	\int \frac{d{^d k}}{(2\pi)^{d}} \frac{1}{k^2 (k-p)^2(k+q)^2} 
	V_{ \mu \nu } ( k - p , - k ) \
	V_{ \alpha \beta  } ( k +q, p - k ) \
	V_{ \rho \sigma  } ( k + q , - k ) ,
\\ \nonumber
\\ \notag
	T^{[4]}_{\mu\nu\alpha\beta\rho\sigma}(p,q) 
& =               \nonumber
\begin{tikzpicture}[baseline={([yshift=-.5ex]current bounding box.center)},thick]
\draw[gluon](1,0)--(2,0); 
\draw (1.8,-0.1) node[below]{$\mu \nu$};
\draw (1.5, 0.2) node[above]{$q$};
\draw [->] (1.25, 0.2) -- (1.75, 0.2);
\draw (1,0) node[left]{$\rho \sigma$};
        \draw (2,0) circle [radius=0.17];
        \draw (2-0.12, - 0.12) -- (2+0.12, 0.12);
        \draw (2+0.12,  -0.12) -- (2-0.12,  0.12);
\draw [particle] (2,0) arc [radius=0.6, start angle=180, delta angle= -180];
\draw [particle] (3.2,0) arc [radius=0.6, start angle=0, delta angle= -180];
\draw (4.2,0.7) node[right]{$ $};
\draw (4.2,0) node[right]{$\alpha \beta$};
\draw (2.5, 0.7) node[above]{$k -p$};
\draw (2.4, - 0.65) node[below]{$k$};
\draw (3.8, 0.25) node[above]{$p$};
\draw [<-] (3.5,0.2) -- (4,0.2);
\draw[gluon](4.2,0)--(3.2,0); 
\end{tikzpicture}
\\
& =
	\int \frac{d{^d k}}{(2\pi)^{d}} \frac{1}{k^2 (k-p)^2} 
	V^{(1)}_{\mu\nu;\rho\sigma}  (k,p-k,q) \
	V_{ \alpha \beta  } ( k , p - k ).  \label{T4}
\end{align}
The vertex function $V_{\mu\nu}(k,\ell)$ was already defined in (\ref{Vmunu});
the remaining ones are
\begin{align}
\begin{tikzpicture}[baseline={([yshift=-.5ex]current bounding box.center)}, thick]
\draw[gluon](-1,1)--(0,0);
\draw [->] (-0.6, 0.9) -- (-0.3,0.65);
\draw[gluon](-1,-1)--(0,0);
\draw [->] (-0.6, -0.9) -- (-0.3,-0.65);
\draw[particle](1,1,0)--(0,0); 
\draw[particle](1,-1,0)--(0,0);
\draw (-1,1) node[left]{$\alpha \beta$};
\draw (-1,-1) node[left]{$\rho \sigma$};
\draw (-0.2,0.7) node[above]{$p$};
\draw (-0.25,-0.7) node[below]{$q$};
\draw (0.35,0.55) node[above]{$k$};
\draw (0.6,-0.5) node[above]{$\ell$};
\end{tikzpicture}
=
\left. W_{ \alpha\beta\rho\sigma }  (k,\ell,p,q) \right|_{p+q+k+\ell = 0}
=
	W^{(1)}_{\alpha\beta\rho\sigma}  (k,\ell)
	+ \xi W^{(2)}_{\alpha\beta\rho\sigma}  (p,q)
\end{align}
where 
\begin{align}
	W^{(1)}_{\alpha\beta\rho\sigma}  (k,\ell)
& =
	- \frac{1}{4} \eta_{\rho (\alpha} \eta_{\beta)\sigma} k\ell
	+ \frac{1}{8}  \eta_{\alpha\beta} \eta_{\rho\sigma} k\ell
	- \frac{1}{4} \eta_{ \alpha \beta } k_{(\rho} \ell_{\sigma)}
	- \frac{1}{4} \eta_{ \rho \sigma } k_{(\alpha} \ell_{\beta)}
	+ \frac{1}{2} k_{(\alpha} \eta_{\beta) ( \rho } \ell_{ \sigma) }
	+ \frac{1}{2} \ell_{(\alpha} \eta_{\beta) ( \rho } k_{ \sigma) }	
\\
	W^{(2)}_{\alpha\beta\rho\sigma}  (p,q)
& =
	\frac{1}{4} \eta_{\alpha\beta} q_\rho q_\sigma
	+ \frac{1}{4}  \eta_{\rho\sigma} p_\alpha p_\beta
	- \frac{1}{4}  \eta_{\alpha\beta} \eta_{\rho\sigma}  q^2
	- \frac{1}{4}  \eta_{\alpha\beta} \eta_{\rho\sigma}  p^2
	+ \frac{3}{4} \eta_{\rho(\alpha} \eta_{\beta)\sigma} pq
	- \frac{1}{2} q_{(\alpha} \eta_{\beta) ( \rho } p_{ \sigma) } 
\\ \nonumber
& \hspace{2em} {}	
	+ \frac{1}{2}  \eta_{\rho (\alpha} \eta_{\beta)\sigma}  q^2
	+ \frac{1}{2}  \eta_{\rho (\alpha} \eta_{\beta)\sigma} p^2
	+ \frac{1}{2}  \eta_{ \rho \sigma } q_\alpha q_\beta
	+ \frac{1}{2}  \eta_{\alpha\beta} p_\rho p_\sigma
	-  q_{(\alpha} \eta_{\beta) ( \rho } q_{ \sigma) }	
	-  p_{(\alpha} \eta_{\beta) ( \rho } p_{ \sigma) }
\\ \nonumber
& \hspace{2em} {}	
	+ \frac{1}{2} \eta_{ \alpha \beta } p_{(\rho} q_{\sigma)}
	+ \frac{1}{2} \eta_{ \rho \sigma } p_{(\alpha} q_{\beta)}
	- \frac{1}{4}  \eta_{\alpha\beta} \eta_{\rho\sigma} pq
	- p_{(\alpha} \eta_{\beta) ( \rho } q_{ \sigma) }	
\end{align}
and
\begin{align}
\begin{tikzpicture}[baseline={([yshift=-.5ex]current bounding box.center)},thick]
\draw[gluon](-1.5,0)--(0,0);
        \draw (0,0) circle [radius=0.17];
        \draw (-0.12, - 0.12) -- (0.12, 0.12);
        \draw (0.12,  -0.12) -- (-0.12,  0.12);
\draw[particle](1,1,0)--(0,0); 
\draw[particle](1,-1,0)--(0,0);
\draw (0,-0.45) node{$\mu \nu$};
\draw (-1.5,0) node[left]{$\rho \sigma$};
\draw (0.35,0.55) node[above]{$k$};
\draw (0.6,-0.5) node[above]{$\ell$};
\draw (-0.75,0.2) node[above]{$q$};
\draw [->] (-1.1, 0.2) -- (-0.6, 0.2);
\end{tikzpicture}
=
V^{(1)}_{\mu\nu;\rho\sigma}  (k,\ell,q)
=
	V^{(1);0}_{\mu\nu;\rho\sigma}  (k,\ell)
	+ V^{(1);1}_{\mu\nu;\rho\sigma}(k+\ell,q)
	+ V^{(1);2}_{\mu\nu;\rho\sigma}(q)
\end{align}
where
\begin{align} \label{V10}
	V^{(1);0}_{\mu\nu ; \rho\sigma}  (k,\ell)
&=
	\frac{1}{2} \eta_{\mu (\rho} \eta_{\sigma)\nu} k \cdot \ell 
	- \frac{1}{2} \eta_{\mu\nu} k_{(\rho} \ell_{\sigma)} 
	- \xi \left( \eta_{\mu (\rho} \eta_{\sigma)\nu} (k+\ell)^2-\eta_{\mu\nu} (k+\ell)_{(\rho}(k+\ell)_{\sigma )} \right)
\\ \label{V11}
	V^{(1);1}_{\mu\nu;\rho\sigma}(\ell,q)
& =
	- \xi\left[  q_{(\mu} \eta_{\nu)(\rho}\ell_{\sigma)}
	- \eta_{\mu\nu} q_{(\rho} \ell_{\sigma)} 
	  - \frac{1}{2} \eta_{\mu (\rho} \eta_{\sigma) \nu} q \cdot \ell
	+ \frac{1}{2} \eta_{\mu\nu} \eta_{\rho\sigma} q \cdot \ell \right]
\\ \label{V12}
	V^{(1);2}_{\mu\nu;\rho\sigma}(q)
& =
	- \xi
	\left[ 
		q_{(\rho}\eta_{\sigma)(\mu} q_{\nu)} 
		- \frac{1}{2} q^2 \eta_{\mu (\rho}\eta_{\sigma)\nu} 
		-  \frac{1}{2} \eta_{ \rho \sigma} q_\mu  q_\nu  
		-  \frac{1}{2} \eta_{ \mu \nu } q_\rho  q_\sigma 
		+ \frac{1}{2} \eta_{\mu\nu} \eta_{\rho\sigma} q^2
	\right]
\end{align}

At second order in $h$ covariant conservation  of the stress tensor requires
\begin{align}\label{covDT}
	\nabla^\mu \braket{ T_{\mu\nu}(x) }
&=
\partial^\mu \braket{ T_{\mu\nu}(x) }_{ \cO (h^2) }\nonumber
\\
& \hspace{2em}
- h^{\mu \rho} \partial_\rho \braket{ T_{\mu\nu}(x) }_{ \cO (h) }
- \frac{1}{2} \left( 2 \partial_\mu h^{\mu\rho} - \partial^\rho h \right) \braket{ T_{\rho\nu}(x) }_{\cO (h) }
- \frac{1}{2} \partial_\nu h^{\mu\rho}  \braket{ T_{ \mu \rho }(x) }_{ \cO (h) }
\end{align}
as can be  confirmed by a somewhat tedious calculation which is, however, completely analogous to the
one performed in \cite{GN}.

To determine the anomaly we recall the known result for the Weyl invariant 
case ($\xi = \frac16$), which reads \cite{Dowker, Brown}
\begin{align}
\mathcal{A}
=	
	g^{(4)\mu\nu}(x) \left\langle T_{\mu\nu}(x) \right\rangle\Big|_{\xi=\frac16}
		\, &= \, \frac{1}{180(4\pi)^2}\left[
		\mathrm{Riem}^2		
		- \mathrm{Ric}^2
		+ \square R
	\right] \\[2mm]
 &= \; 
	\frac{1}{180(4\pi)^2}\left[ -\frac12 {\bf E}_4 + \frac32 C^2 + \Box R \right].
\end{align}

We now perform the calculation for arbitrary $\xi$. In this case the computation is 
substantially more involved, and for this reason we had to make use of a 
Mathematica code, in particular we exploited the HEPMath package \cite{hepmath}. Schematically for the two contributions to (\ref{A}) we find
\begin{align}\label{cac}
	\ g^{(4)\mu\nu}(x) \big\langle T_{\mu\nu}(x) \big\rangle_\xi 
& \,=\, 
	- \frac{ (6\xi -1)^2 }{12(4\pi)^2 \ \varepsilon}\square R  \,+\, A \,+\, \cO(\varepsilon) \notag
\\
	\left\langle  g^{\mu\nu}(x) T_{\mu\nu}(x) \right\rangle_\xi
& \,=\, 
	- \frac{ (6\xi -1)^2 }{12(4\pi)^2 \ \varepsilon}\square R  \,+\, B \,+\, \cO(\varepsilon)
\end{align}
for the regularized expressions.
The poles correctly cancel with each other and vanish, as does $B$, when $\xi = 1/6$.
However, for generic $\xi$ the functions $A$ and $B$ are very complicated 
with about 15\,000 terms each; most of these are non-local, involving expressions
like $1/((pq)^2-p^2 q^2)^4$, $\log p^2$, $\log{} (p+q)^2$ in the momentum space integrals.
All these terms come from the diagrams with three external legs, 
as well as the finite scalar loop integral $J_{111}$ (see ~\cite{GN}).
Remarkably in the difference $A-B$, all these unwanted terms cancel, 
leaving a  much simpler expression that in momentum space contains 
less than 200 terms and combines correctly into the second order expressions
required for the covariant expressions in the curvature tensor. The final result is
\begin{align}
\mathcal{A}_\xi
& =	
	g^{(4)\mu\nu}(x) \left\langle T_{\mu\nu}(x) \right\rangle_\xi
	- \left\langle  g^{\mu\nu}(x) T_{\mu\nu}(x) \right\rangle_\xi \notag
\\[2mm]
& =
	\frac{1}{180(4\pi)^2}\left[
		\mathrm{Riem}^2		
		- \mathrm{Ric}^2
		+ \left( 1 - 10(1-6 \xi)^2  \right)  \square R
		+ \frac{5}{2} (1-6\xi)^2 R^2
	\right]\label{caa}
\end{align}
notice that the coefficient in front of $\square R$ matches the result from $\cO(h)$ in \eqref{anxiD}. 
This result matches with the one reported in \cite{BD} up to the coefficient of $\square R$, which matches the one we found in \eqref{anxi4}. Following the discussion around equations \eqref{anxiD} and \eqref{anxi4}, we can trace the difference 
to the subtraction of a different classical contribution, which is not made explicit in \cite{BD}.
Observe also that, as we already anticipated, for $\xi\neq \frac16$ 
there appears on the r.h.s. a contribution $\propto R^2$ which does not satisfy the WZ condition.
The result shows that the anomaly proper -- that is, the terms that cannot be removed
by local counterterms and that satisfy the WZ condition --
are indeed independent of $\xi$  and thus universal. However it remains to be
seen whether this conclusion also holds in a more general context.

As for $\cO(h)$ we can now explicitly exhibit the structure of the
pole of $\langle T_{\mu\nu} \rangle$ at order $\cO(h^2)$.
The pole of the expectation value of the stress-energy tensor is a local generally covariant expression with four derivatives acting on the metric. This constrains the expression to 
be of the form
\begin{equation}\label{cab}
\begin{aligned}
	\langle T_{\mu\nu} \rangle
 =
	\frac{1}{(4\pi)^2 \varepsilon}
&	\Big[
		a_1\ g_{\mu\nu} R^2
 		+ a_2\ R R_{\mu\nu}
		+ a_3\ g_{\mu\nu} R_{\alpha\beta} R^{\alpha\beta}
		+ a_4\ R_{\mu}^\alpha R_{\alpha\nu}
		+ a_5\ R^{\alpha\beta} R_{\mu\alpha\beta\nu}	
\\
& \hspace{0.5em}		
		+ a_6\ R\indices{_\mu^{\alpha\beta\gamma}} R_{\nu\alpha\beta\gamma}
		+ a_7\ g_{\mu\nu}   R_{\alpha\beta\gamma\delta}   R^{\alpha\beta\gamma\delta}
		+ a_8\ \nabla_\mu \nabla_\nu R
		+ a_9\ g_{\mu\nu} \square R
		+ a_{10}\ \square R_{\mu\nu}
	\Big],
\end{aligned}
\end{equation}
Any other term can be related to those written above via Bianchi identities 
and symmetry arguments.
Writing out the $\cO(h^2)$ expansions for all these contributions, and matching with the 
second order results of our computations we get
\begin{gather} 
  a_1 =  \frac{(1-6\xi)^2}{144}\;, \qquad   a_2= - \frac{(1-6\xi)^2}{36}\;, \qquad  
  a_3=  - \frac1{360}\;,    \notag \\[2mm]
a_4 = \frac1{45}\;, \qquad  a_5= \frac1{90}\;, \qquad a_6 = - \frac1{90}\;, \qquad 
a_7= \frac1{360}\;, \notag \\[2mm]
a_8 = \frac{1 - 10 \xi + 30 \xi^2}{30}\;, \qquad 
a_9= - \frac{3 - 40 \xi + 120 \xi^2}{120}, \qquad a_{10}= - \frac1{60}.\label{coeff}
\end{gather}
The coefficients $a_8$, $a_9$, $a_{10}$ match those computed at order 
$\cO(h)$ (as they should), and therefore considering the trace we recover also \eqref{cac}.
It is also noteworthy that, since $g^{(4)\, \mu\nu} \braket{T_{\mu\nu}} \sim \square R / \varepsilon$, 
it follows that 
\begin{align}
4a_1 + a_2 &=0 & 
4a_3 + a_4 -a_5&=0 &
4a_7 + a_6 &=0
\end{align} as they correspond to the coefficients of $R^2$, $\mathrm{Ric}^2$ and 
$\mathrm{Riem}^2$. We can see that the coefficients in (\ref{coeff}) indeed respect 
this constraint, and this is a nontrivial consistency check of the result.
Furthermore, from the general arguments of section \ref{sec:anomaly}, and
more specifically exploiting formula (\ref{AA}), the anomaly is 
\begin{equation}
\begin{aligned}
	\mathcal{A}_{\xi}
 =
	\frac{2}{(4\pi)^2}
&	\Big[
		a_1\  R^2
		+ a_3\  \mathrm{Ric}^2
		+ a_7\   \mathrm{Riem}^2
		+ a_9\  \square R
	\Big],
\end{aligned}
\end{equation}
which indeed agrees with expression \eqref{caa} upon substituting \eqref{coeff}.

Following the derivation of the conformal anomaly often done in the literature (see \eg \cite{BD} for a complete exposition), we have independently confirmed 
the coefficients \eqref{coeff} by computing the pole of $\braket{T_{\mu\nu}} = - (2/\sqrt{-g}) \delta \Gamma/\delta g^{\mu\nu}$ from the regularised effective action $\Gamma$ computed with a heat kernel expansion. The heat kernel method yields the following explicit expression for the (regularised) effective action:
\begin{equation}
\Gamma[g] = -\frac{1}{2} \log \det(-\Box + \xi R) = \frac{1}{(4\pi)^2 2 \varepsilon} \ \int  \sqrt{-g}\ a_2 + \cO (\varepsilon^0)
\end{equation}
where $a_2(x) $ reads (we are neglecting here a $\Box R$ contribution, as it is a boundary term)
\begin{equation}
a_2(x) = \frac{1}{180} \mathrm{Riem}^2 - \frac{1}{180} \mathrm{Ric}^2 + \frac{1}{72} (1-6\xi)^2 R^2.
\end{equation}
Explicit expressions for the variations $\delta(\sqrt{-g}\, \mathrm{Riem}^2) $, $\delta(\sqrt{-g}\, \mathrm{Ric}^2) $, $\delta(\sqrt{-g}\, R^2) $ that can be usefully employed for this calculation
can be found in \cite{GMN}.

\vspace{5mm} 
\noindent
 {\bf Acknowledgments:} 
 The work of  H.~Nicolai is supported in part by the European Research Council (ERC) under the 
 European Union's Horizon 2020 research and innovation programme (grant agreement 
 No 740209).

\end{document}